\documentclass[aps,prl,twocolumn,floats,showpacs,superscriptaddress]{revtex4}
\usepackage{graphicx,epsfig}
\usepackage{times}
\usepackage{graphics,dcolumn,bm,float}
\usepackage{amssymb,amsmath,rotate,color}

\begin{document}
\unitlength 1 cm
\newcommand{\be}{\begin{equation}}
\newcommand{\ee}{\end{equation}}
\newcommand{\bearr}{\begin{eqnarray}}
\newcommand{\eearr}{\end{eqnarray}}
\newcommand{\nn}{\nonumber}
\newcommand{\vk}{\vec k}
\newcommand{\vp}{\vec p}
\newcommand{\vq}{\vec q}
\newcommand{\vkp}{\vec {k'}}
\newcommand{\vpp}{\vec {p'}}
\newcommand{\vqp}{\vec {q'}}
\newcommand{\bk}{{\bf k}}
\newcommand{\bp}{{\bf p}}
\newcommand{\bq}{{\bf q}}
\newcommand{\br}{{\bf r}}
\newcommand{\bR}{{\bf R}}
\newcommand{\up}{\uparrow}
\newcommand{\down}{\downarrow}
\newcommand{\fns}{\footnotesize}
\newcommand{\ns}{\normalsize}
\newcommand{\cdag}{c^{\dagger}}

\title{Flow Equations for the Ionic Hubbard Model}
\author{Mohsen Hafez}
\affiliation{Department of Physics, Tarbiat Modares University, Tehran, Iran}

\author{S. A. Jafari{\footnote {Electronic address:
akbar.jafari@gmail.com}}}

\affiliation{Department of Physics, Isfahan University of
Technology, Isfahan 84154-83111, Iran}
\affiliation{The Abdus Salam ICTP, 34100 Trieste, Italy}

\author{M. R. Abolhassani}
\affiliation{Department of Physics, Tarbiat Modares University,
Tehran, Iran}

\begin{abstract}
  Taking the site-diagonal terms of the one-dimensional ionic Hubbard model (IHM)
as $H_0$, we employ Continuous Unitary Transformations
(CUT) to obtain a "classical" effective Hamiltonian in which
hopping term has been integrated out. For this Hamiltonian spin gap and
charge gap are calculated at half-filling and subject to periodic
boundary conditions. Our calculations indicate two transition points.
In fixed $\Delta$, as $U$ increases from zero, there is a region in
which both spin gap and charge gap are positive and identical; characteristic of band
insulators. Upon further increasing $U$, first transition occurs at
$U=U_{c_{1}}$, where spin and charge gaps both vanish and remain zero up
to $U=U_{c_{2}}$. A gap-less state in charge and spin sectors
characterizes a metal. For $U>U_{c_{2}}$ spin gap remains zero and
charge gap becomes positive. This third region corresponds to a Mott
insulator in which charge excitations are gaped, while spin
excitations remain gap-less.
\end{abstract}
\maketitle

\section{Introduction}

  Ionic Hubbard model (IHM) has been used to study neutral-ionic
transition in organic compounds~\cite{Nagaosa} and understanding
role of strong correlations in ferroelectricity of metal oxides such as
BaTiO$_3$, KNbO$_3$, KTaO$_3$~\cite{Egami}, 
as well as some quasi one-dimensional materials such as 
(TaSe$_4$)$_2$I, K$_{0.3}$MoO$_3$~\cite{RefolioJPCM}.
IHM Hamiltonian includes an staggered one-body external potential in addition 
to the Hubbard Hamiltonian. The Hamiltonian for this model is as follows:
\bearr
  H &=&-t\sum_{i\sigma}(c^{\dag}_{i\sigma}c_{i+1\sigma}+h.c.)+
  U\sum_{i}c^{\dag}_{i\uparrow}c^{\dag}_{i\downarrow}c_{i\downarrow}c_{i\uparrow}\nn\\
  &&+\frac{\Delta}{2}\sum_{i\sigma}(-1)^{i}c^{\dag}_{i\sigma}c_{i\sigma},
  \label{ihm.eqn}
\eearr
where $c_{i\sigma}$ ($c^{\dag}_{i\sigma}$) is the usual annihilation (creation) operator
at site $i$ with spin $\sigma$, $t$ the nearest neighbor hopping amplitude,
$U$ the on-site coulomb interaction parameter,
and $\Delta$ a one-body ionic potential. This model has
a very interesting phase diagram at zero temperature at half-filling.
When $\Delta\ll$ $U$ this Hamiltonian, like the usual Hubbard Hamiltonian,
is transformed to the Heisenberg spin Hamiltonian~\cite{Nagaosa,Fazekas} that
describes a Mott insulator. In the opposite limit $\Delta\gg U$, one can
ignore $U$, and a Bogoliubov transformation gives a
simple band insulator with gap $\Delta$. The limit $U\gg \Delta$
is a many-body insulator, while the limit $U\ll \Delta$ is a 
one-body insulator. Hence this model is a basis to study the 
issues like matrix-element effects in optical spectra of 
band insulators, versus many-body insulators~\cite{TohyamaNLO}.

Apart from offering two entirely different insulating states, yet there remains 
interesting question of the intermediate phase: What is the 
nature of the ground state for $U\sim \Delta$?
Researchers have used various methods such as 
exact diagonalization (ED)~\cite{Egami,Sorella},
density matrix renormalization group (DMRG)~\cite{Manmana,Legeza,Brune}, 
quantum Monte Carlo (QMC)~\cite{Nagaosa,Bouadim,Wilkens,Refolio},
dynamical mean field theory (DMFT)~\cite{Garg,Craco,Kancharla}, etc. to study the properties of this model.

The nature of intermediate phase still remains controversial: 
Bosonization study of Fabrizio and coworkers~\cite{Fabrizio} indicates
an spontaneously dimerized phase, which is also supported by spin-particle
transformation study of Batista and coworkers~\cite{Batista}. There are also
other works supporting this scenario~\cite{Manmana,Legeza,Wilkens,Kancharla}.
On the other hand, there are studies showing that the intermediate phase
is metallic~\cite{Nagaosa,Bouadim,Garg,Sorella}.
Brune finds a metallic transition point in between the band and Mott
insulating states, with simultaneous bond order~\cite{Brune}. Craco et. al. 
in addition to metallic region, report on a coexistence phase between
the band and Mott insulating states~\cite{Craco}.

In this work, we employ the method of continuous unitary transformations
or flow equations to study the half-filled one-dimensional IHM in zero temperature subject
to periodic  boundary conditions. CUT method is used to obtain a
"classical" effective Hamiltonian for IHM. This effective Hamiltonian which is free of
fermionic minus sign problem, provides a simple picture of the nature of excitations when 
interpreted in terms of the transformed ground state. Then for $L=20$ sites
and with $\Delta=20$ (energies are in units of $t$) spin and charge gap are numerically calculated for different
values of $U$. These calculations indicate two transition points. For
$U<U_{c_{1}}$, spin and charge gap are both positive, for
$U_{c_{1}}<U<U_{c_{2}}$ are both zero and for $U>U_{c_{2}}$ spin
gap remains zero while charge gap becomes positive. 
Hence this the transformed effective Hamiltonian gives band insulator
for small $U$ region ($U<U_{c_1}$), Mott insulator for large $U$
regime ($U>U_{c_2}$), and metal for the region in between.

\section{Atomic Limit}
\begin{figure}[t]
  \begin{center}
    \includegraphics[width=7cm]{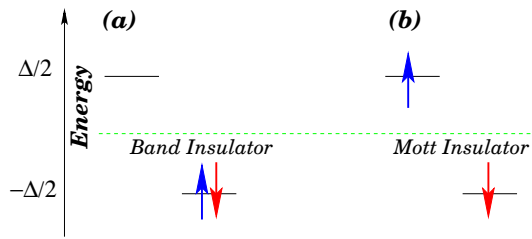}
    \caption{(Color online) Lowest energy configurations for (a) Perfect band
    insulator, (b) Perfect Mott insulator.}
    \label{schematic.fig}
  \end{center}
\end{figure}

In the limits $U\ll\Delta$ and $\Delta\ll U$, the model becomes
easy to understand and gives band and Mott insulators, respectively.
Yet, another interesting limit in which the phase diagram becomes particularly
simple is the atomic limit ($t=0$), 
\be
   E[n_{i\sigma}]=\frac{\Delta}{2}\sum_{i\sigma}
   (-1)^{i}n_{i\sigma} +U \sum_{i} n_{i\uparrow}n_{i\downarrow},
   \label{atomiclimit.eqn}
\ee 
where the Hamiltonian can be written solely in terms of classical 
(commuting) variables $n_{i\sigma}$. In this sense, the atomic limit
corresponds to the classic limit of a quantum problem, and it 
becomes much easier to analyze.
Any configuration $|\{n_{i\sigma}\}\rangle$ is an eigen-state of
this Hamiltonian. The hopping term which causes quantum fluctuations
of various $|\{n_{i\sigma}\}\rangle$ configurations by admixing them
is absent here. Hence the problem reduces to finding out the
configuration with lowest energy. To this end, we consider a unit
cell in real space of the 1D lattice shown in Fig.~\ref{schematic.fig}. 
Since this Hamiltonian contains purely local degrees of freedom,
we consider a zero momentum mode ($k=0$) which corresponds to
repeating this unit cell all over the 1D chain. In part (a), a state
with energy $E_a=U-\Delta$ per unit cell is shown, while the one in
part (b) has energy $E_b=0$ per unit cell. For $E_a<E_b$, i.e.
$U-\Delta < 0$, state (a) indicating a band insulator is stabilized,
while for $U>\Delta$, the Mott insulating state with energy $E_b$ is
stabilized. At $U=\Delta$ these two states are degenerate, and hence
in $k=0$ mode, $\uparrow$ spin can freely move in the unit cell of
Fig.~\ref{schematic.fig}. Therefore the picture in the atomic limit
is as follows: Eq.~(\ref{atomiclimit.eqn}) describes a Mott
insulator for $\Delta<U$ and a band insulator for $\Delta>U$. The
transition line $\Delta=U$ is a metal phase shrinked to a line.

   The question we would like to address in this paper is,
what happens if we turn on the quantum fluctuations around the
Hamiltonian (\ref{atomiclimit.eqn}) by introducing a finite hopping
$t$? It turns out that CUT is a very suitable tool to investigate
this {\em strong coupling} problem. 
We advertise the result in advance: The inclusion of the hopping $t$ will renormalize
$U\to\tilde U(t,U,\Delta), \Delta\to\tilde\Delta(t,U,\Delta),t\to \tilde t=0$, 
as well as generating new couplings.

\section{Continuous Unitary Transformations}
Flow equations approach or, CUT method~\cite{KehreinBook} was introduced independently 
by Glazek and Wilson~\cite{Wilson}, and  Wegner~\cite{Wegner}. Since then,
this method has been applied extensively to study various models in
condensed matter and high energy physics. Some examples include, the Anderson model~\cite{Kehrein},
Lipkin model~\cite{Stein}, mapping of the electron-phonon interaction to an
effective electron-electron interaction~\cite{Lenz}, Sine-Gordon
model~\cite{Sine}, RKKY interaction~\cite{Stein99}, as well as 
non-equilibrium dynamics of strongly correlated systems~\cite{Moeckel} 
and an attempt for a unified description of Fermi and Luttinger liquids
in all energy scales~\cite{Heidbrink}.

In this method the Hamiltonian is transformed by
a unitary operator $U(\ell)$ where $\ell$ is a parameter between
zero to infinity. Flow equation for $H(\ell)=U^{\dag}(\ell)HU(\ell)$ becomes,
\be
   \partial_{\ell}H(\ell)=[\eta(\ell),H(\ell)],
   \label{cut.eqn}
\ee
where $\eta(\ell)=\partial_{\ell}U^{\dag}(\ell).U(\ell)$ is an anti-Hermitian
operator called generator. Here $U(\ell=0)=1$, as at $\ell=0$
transformed Hamiltonian is equal to $H$. Wegner suggested the following
generator~\cite{Wegner}:
\be
   \eta(\ell)=[H^{d}(\ell),H^{r}(\ell)],
   \label{wegner.eqn}
\ee
where $H^{d}(\ell)$ $\left(H^{r}(\ell)\right)$ is diagonal (off-diagonal) part of the 
Hamiltonian. With this choice for the generator, transformed Hamiltonian 
flows towards a diagonal or block-diagonal form in the $\ell\to\infty$ limit.
Mielke introduced another generator that is specially useful for band matrices~\cite{Mielke}
and preserves the band nature of the Hamiltonian matrix.
In this section, we use Wegner generator to obtain an effective Hamiltonian for IHM.

It is obvious from Eq.~(\ref{cut.eqn}) that the flow equations
without any approximation is very hard to solve for Hamiltonians
that include interaction. In applying the flow equations to
IHM we use an approximation similar to the one used by Kehrein and
Mielke in the context of Anderson model~\cite{Kehrein}. We
approximate $H(\ell)$  as,
\bearr
   H(\ell)\!\!&=&\!\!-t(\ell)\!\sum_{i\sigma}(c^{\dag}_{i\sigma}c_{i+1\sigma}+h.c.)\!+
   \!\frac{\Delta(\ell)}{2}\!\sum_{i\sigma}(-1)^{i}c^{\dag}_{i\sigma}c_{i\sigma}\nn\\&&+
   \frac{U(\ell)}{2}\sum_{i\sigma\sigma\prime}
   c^{\dag}_{i\sigma}c^{\dag}_{i\sigma\prime}c_{i\sigma\prime}c_{i\sigma}\nn\\&&+
   V(\ell)\sum_{i\sigma\sigma\prime}c^{\dag}_{i\sigma}c^{\dag}_{i+1\sigma\prime}
   c_{i+1\sigma\prime}c_{i\sigma}.
   \label{IHMell.eqn}
\eearr
with initial conditions $t(0)=1$, $\Delta(0)=\Delta$, $U(0)=U$ and
$V(0)=0$. Physical motivation behind considering n.n. Coulomb interaction
$V$ is as follows~\cite{Legeza}: In the simple phase diagram shown in Fig.~\ref{schematic.fig}
for a fixed $\Delta$, as one increases $U$, the ionicity $n_B-n_A$ which is
the population difference between $A,B$ lattice sites, suddenly drops
from $1$ to $0$ at $U_{c_2}=\Delta$. If an attractive n.n. term $V$
can be generated, it opens up an intermediate phase in which
the $0<n_B-n_A<1$; i.e. the unit cell gently looses its polarization
with respect to the half-filled background and turns into a Mott insulator.

In Eq.~(\ref{IHMell.eqn}), the first term is taken to be 
off diagonal and other terms are diagonal. This choice resembles
the strong coupling perturbation schemes starting from the atomic limit.
There one usually has to organize the hopping processes into those terms
which change the double occupancy, and those which do not~\cite{Fazekas}.
But here we show that the inclusion of smallest amount of $\Delta$
makes it possible to renormalize the whole hopping processes to zero
(Eq.~\ref{19}).

Using equation (\ref{wegner.eqn}),
With this convention, the Wegner generator is obtained as: \bearr
   \!\!\!\!&&\eta(\ell)=t(\ell)\Delta(\ell)\sum_{i\sigma}(-1)^{i}(c^{\dag}_{i+1\sigma}c_{i\sigma}-h.c.)\nn\\
   \!\!\!\!&&-t(\ell)U(\ell)\sum_{i\sigma\sigma\prime}(c^{\dag}_{i\sigma}
   c^{\dag}_{i\sigma\prime}c_{i\sigma\prime}c_{i+1\sigma}
   -c^{\dag}_{i-1\sigma}c^{\dag}_{i\sigma\prime}c_{i\sigma\prime}c_{i\sigma}-h.c.)\nn\\
   \!\!\!\!&&-t(\ell)V(\ell)\sum_{i\sigma\sigma\prime}(c^{\dag}_{i\sigma}
   c^{\dag}_{i+1\sigma\prime}c_{i+1\sigma\prime}c_{i+1\sigma}
   +c^{\dag}_{i\sigma}c^{\dag}_{i+1\sigma\prime}c_{i+2\sigma\prime}c_{i\sigma}\nn\\
   \!\!\!\!&&-c^{\dag}_{i\sigma}c^{\dag}_{i\sigma\prime}c_{i+1\sigma\prime}c_{i\sigma}
   -c^{\dag}_{i-1\sigma}c^{\dag}_{i+1\sigma\prime}c_{i+1\sigma\prime}c_{i\sigma}-h.c.)
   \label{etaIHM.eqn}
\eearr To obtain the flow equations, we need to calculate the
commutator of $\eta(\ell)$ and $H(\ell)$. To do so, we split
$\eta(\ell)$ into three parts \be
   \eta(\ell)=\eta_{1}(\ell)+\eta_{2}(\ell)+\eta_{3}(\ell),
\ee
where $\eta_{1}(\ell)$, $\eta_{2}(\ell)$, and $\eta_{3}(\ell)$ denote
the first, second, and third terms in equation (\ref{etaIHM.eqn}) respectively.
Similarly, $H(\ell)$ in equation~(\ref{IHMell.eqn}) has four terms
\be
  H(\ell)=H_{1}(\ell)+H_{2}(\ell)+H_{3}(\ell)+H_{4}(\ell),
\ee
where $H_{1}(\ell)\ldots H_{4}(\ell)$ correspond to four terms in Eq.~(\ref{IHMell.eqn}).
Therefore:
\bearr
   [\eta(\ell),H(\ell)]&=&[\eta_{1}(\ell),H_{1}(\ell)+H_{2}(\ell)]+[\eta_{2}(\ell),H_{1}(\ell)]\nn\\
   &&+[\eta_{3}(\ell),H_{1}(\ell)]+ \mbox{irrelevant terms},
   \label{commutator.eqn}
\eearr
where "irrelevant terms" are newly generated couplings which are not similar to Eq.~(\ref{IHMell.eqn}).
The commutators are calculated as follows:
\bearr
   &&[\eta(\ell),H_{1}(\ell)+H_{2}(\ell)]=\nn\\
   &&2t^{2}(\ell)\Delta(\ell)\sum_{i\sigma}(-1)^{i}
   (c^{\dag}_{i\sigma}c_{i\sigma}+c^{\dag}_{i\sigma}c_{i+2\sigma}+h.c.)\nn\\
   &&+t(\ell)\Delta^{2}(\ell)\sum_{i\sigma}(c^{\dag}_{i+1\sigma}c_{i\sigma}+h.c.)
   \label{etaH1H2.eqn},
\eearr
and
\bearr
   &[\eta_{2}(\ell),H_{1}(\ell)]=t^{2}(\ell)U(\ell)\sum_{i\sigma\sigma'}\sum_{j\beta}\times \nn\\
   &([c^{\dag}_{i\sigma}c^{\dag}_{i\sigma'}c_{i\sigma'}c_{i+1\sigma},c^{\dag}_{j\beta}c_{j+1\beta}]+
   [c^{\dag}_{i\sigma}c^{\dag}_{i\sigma'}c_{i\sigma'}c_{i+1\sigma},
   c^{\dag}_{j+1\beta}c_{j\beta}]\nn\\
   &-[c^{\dag}_{i-1\sigma}c^{\dag}_{i\sigma'}c_{i\sigma'}c_{i\sigma},c^{\dag}_{j\beta}c_{j+1\beta}]
   -[c^{\dag}_{i-1\sigma}c^{\dag}_{i\sigma'}c_{i\sigma'}c_{i\sigma},c^{\dag}_{j+1\beta}c_{j\beta}]\nn\\
   &+h.c.).
   \label{18-7-b}
\eearr
First and third terms are irrelevant,while second and fourth terms respectively become
\bearr
   &&\left(c^{\dag}_{i\sigma}c^{\dag}_{i\sigma'}c_{i\sigma'}c_{i\sigma}\delta_{i,j}\delta_{\sigma\beta}
   -c^{\dag}_{i+1\sigma}c^{\dag}_{i\sigma'}c_{i\sigma'}c_{i+1\sigma}\delta_{i,j}\delta_{\sigma\beta}\right)-\nn\\
   &&\left(c^{\dag}_{i-1\sigma}c^{\dag}_{i\sigma'}c_{i\sigma'}c_{i-1\sigma}\delta_{i,j+1}\delta_{\sigma\beta}
   -c^{\dag}_{i\sigma}c^{\dag}_{i\sigma'}c_{i\sigma'}c_{i\sigma}\delta_{i-1,j}\delta_{\sigma\beta}\right)\nn\\
   &&+\mbox{irrelevant terms}.\nn
\eearr 
Substituting in Eq.~(\ref {18-7-b}) and dropping irrelevant terms gives, \bearr
   &&[\eta_{2}(\ell),H_{1}(\ell)]=2t^{2}(\ell)U(\ell)
   \label{eta2H1.eqn}\\
   &&\times\sum_{i\sigma\sigma'}(c^{\dag}_{i\sigma}c^{\dag}_{i\sigma'}c_{i\sigma'}c_{i\sigma}
   -c^{\dag}_{i+1\sigma}c^{\dag}_{i\sigma'}c_{i\sigma'}c_{i+1\sigma}+h.c.)\nn
\eearr
The last commutator up to irrelevant terms becomes:
\bearr
   &&[\eta_{3}(\ell),H_{1}(\ell)]=2t^{2}(\ell)V(\ell)   \label{eta3H1.eqn} \\
   &&\times\sum_{i\sigma\sigma'}(2c^{\dag}_{i\sigma}c^{\dag}_{i+1\sigma'}c_{i+1\sigma'}c_{i\sigma}
   -c^{\dag}_{i\sigma}c^{\dag}_{i\sigma'}c_{i\sigma'}c_{i\sigma}+h.c.)\nn
\eearr

Substitution of Eqs.~(\ref{etaH1H2.eqn}), (\ref{eta2H1.eqn}) and (\ref{eta3H1.eqn}) in
Eq.~(\ref{commutator.eqn}), gives the following set of flow equations
\bearr
   &&\partial_{\ell}t(\ell)=-t(\ell)\Delta^{2}(\ell)\label{19}\\
   &&\partial_{\ell}\Delta(\ell)=8t^{2}(\ell)\Delta(\ell)\label{20}\\
   &&\partial_{\ell}U(\ell)=8t^{2}(\ell)\left(U(\ell)-V(\ell)\right)\label{21}\\
   &&\partial_{\ell}V(\ell)=4t^{2}\left(\ell)(2V(\ell)-U(\ell)\right)\label{22}
\eearr
We obtain a closed form solution for the above set or equations at
$\ell\rightarrow\infty$.
\bearr
   &t(\infty)=0 \label{23}\\
   &\Delta(\infty)=(8+\Delta^{2})^{\frac{1}{2}}\label{24}\\
   &U(\infty)=\frac{U}{2}\frac{(8+\Delta^{2})^{\frac{1}{2}}((8+\Delta^{2})^{\frac{\sqrt{2}}{4}}+
   (8+\Delta^{2})^{\frac{-\sqrt{2}}{4}}\Delta^{\sqrt{2}})}{\Delta^{1+\frac{\sqrt{2}}{2}}}\label{25}\\
   &V(\infty)=\frac{\sqrt{2}U}{4}\frac{(8+\Delta^{2})^{\frac{1}{2}}(-(8+\Delta^{2})^{\frac{\sqrt{2}}{4}}+
  (8+\Delta^{2})^{\frac{-\sqrt{2}}{4}}\Delta^{\sqrt{2}})}{\Delta^{1+\frac{\sqrt{2}}{2}}}\label{26}
\eearr
Eq.~(\ref{23}) shows that the Hamiltonian is diagonal in infinity.
The hopping term has been renormalized to zero. This has become
possible because of the existence of a non-zero value of $\Delta(\ell=0)$ (Eq.~\ref{19}).
Presence of any non-zero $\Delta$ changes the symmetry of the problem, and
also introduces a new energy scales which makes the life different from that of the
usual Mott-Hubbard physics. In case of a pure Hubbard model, one needs to
take split the hopping terms into those which change the double occupancy, and
those which do not~\cite{Uhrig-t-J,KehreinDiscussion}. But here due to 
presence of a new energy scale, $\Delta$, the whole hopping term can be renormalized
to zero.

Eqs.~(\ref{24}), (\ref{25}), and (\ref{26}) indicate that the result of
integrating out the hopping term in the Hamiltonian renormalizes
the diagonal terms. Integrating out
the hopping term also induces new coupling which are not of the
form (\ref{IHMell.eqn}), and hence irrelevant.
Therefore the fixed point of the one dimensional IHM is described by the effective
Hamiltonian
\bearr
   H(\infty)&=&\frac{\Delta(\infty)}{2}\sum_{i\sigma}(-1)^{i}n_{i\sigma}+
   U(\infty)\sum_{i}n_{i\uparrow}n_{i\downarrow}\nn\\
   &&+ V(\infty)\sum_{i\sigma\sigma\prime}n_{i\sigma}n_{i+1\sigma\prime},
   \label{Heff.eqn}
\eearr
where $n_{i\sigma}$ is the number of electron in site $i$ with spin $\sigma$.
Also note from Eq.~(\ref{26}) that for any
$U,\Delta>0$, $V(\infty)$ is always negative. Hence the induced n.n. Coulomb interaction
is attractive. As a consistency check we note that our effective Hamiltonian 
in the limit $U,\Delta\gg1$ Eq.~(\ref{Heff.eqn}) reduces to
the atomic limit ($t=0$) of Eq.~(\ref{ihm.eqn}).

\section{Phase transitions}
\begin{figure}[t]
  \begin{center}
    \vspace{0.0 cm}
    \includegraphics[scale=0.55,angle=-90]{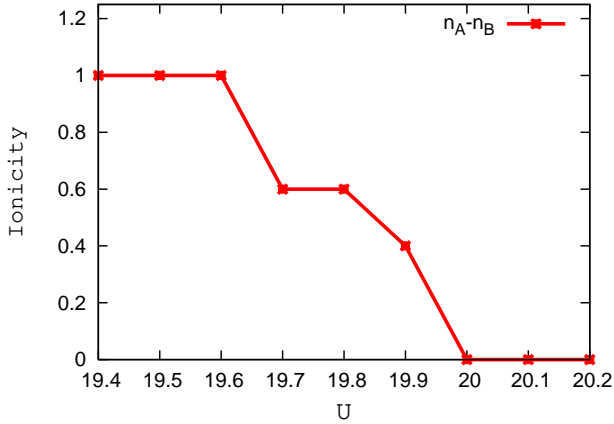}
    \caption{Ionicity for $\Delta=20$, $t=1$ and $L=20$ sites.
    }
    \label{ionicity.fig}
  \end{center}
\end{figure}
\subsection{Ionicity}
Fig.~\ref{schematic.fig} represents two limiting case of the
Hamiltonian~(\ref{atomiclimit.eqn}). For the band insulator,
where both electrons of spin $\uparrow$ and $\downarrow$ prefer
the B sites with energy $-\Delta/2$, the unit cell is fully polarized.
In the opposite limit of the Mott insulator, there is only one electron
per site, and the unit cell polarization is zero. Therefore we
define the ionicity for IHM as $n_{B}-n_{A}$ where  $n_{B}$ and $n_{A}$ are
the density of electron in B and A sites respectively:
\bearr
   && n_{B}=\frac{1}{N}\sum_{\sigma,i\in {\rm B}}\langle\widehat{n}_{i\sigma}\rangle \label{30} \\
   && n_{A}=\frac{1}{N}\sum_{\sigma,i\in {\rm A}}\langle\widehat{n}_{i\sigma}\rangle \label{31}
\eearr

The ionicity for the IHM is depicted in  Fig.~\ref{ionicity.fig}.
As can be seen in the figure, the ionicity is
1 for $U<U_{c_1}$, indicating a band insulating state. For
$U>U_{c_2}$, the ionicity becomes zero, characterizing a Mott
insulator. The intermediate region $U_{c_1}<U<U_{c_2}$ is
characterized by $1>n_B-n_A>0$. This basic physics in contained in
{\em attractive} nature of the induced $V$ term in
Eq.~(\ref{Heff.eqn}). Without this term, there is only one phase
transition at $U_c=\Delta$ (Eq.~\ref{atomiclimit.eqn}). However, the
renormalization process leading to generation of $V$ term, opens up
an intermediate state allowing for the smooth decrease of the
ionicity from $1$ (Band insulator) to $0$ (Mott insulator), by
partially polarizing the unit cell.

  In this picture the reduction in the degree of polarization of the
unit cell to intermediate values facilitates the charge transfer
between the two sites of each unit cell. Therefore the resulting intermediate
state must be a metal.

\subsection{Spin and charge gaps}
\begin{figure}[t]
  \begin{center}
    \vspace{0.0 cm}
    \includegraphics[scale=0.55,angle=-90]{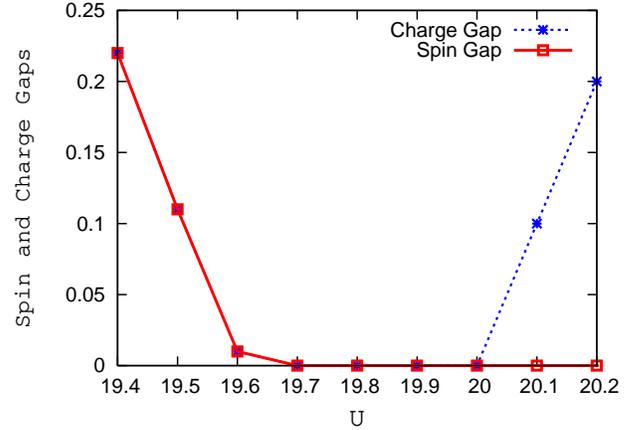}
    \caption{Spin and charge gaps versus U for $\Delta=20$, $t=1$ and $L=20$ sites.
    }
    \label{Gap.fig}
  \end{center}
\end{figure}
To investigate the intermediate state further, we numerically calculate the
charge and spin gaps for the IHM. Spin and charge gaps are defined
as follows~\cite{Dziurzik}    
\bearr
  &\Delta_{s}=E_{0}(\frac{N}{2}+1,\frac{N}{2}-1)-E_{0}(\frac{N}{2},\frac{N}{2}),\label{28}\\
  &\Delta_{c}=\frac{E_{0}(\frac{N}{2}+1,\frac{N}{2}+1)+
   E_{0}(\frac{N}{2}-1,\frac{N}{2}-1)}{2}-E_{0}(\frac{N}{2},\frac{N}{2}), \label{29}
\eearr
where $E_{0}(N_{\uparrow},N_{\downarrow})$ is the ground state energy
in a sector with $N_{\uparrow}$ ($N_{\downarrow}$) electron in spin up (down) state.
Since the unitary
transformations do not affect the level spacing, the
spin and charge gaps defined above will be the same for effective
Hamiltonian (\ref{Heff.eqn}) as well as the original Hamiltonian (\ref{ihm.eqn}).
Fortunately the effective Hamiltonian~(\ref{Heff.eqn}) is classical and hence
free of fermionic minus sign issues. Therefore the ground state energy
$E_{0}(N_{\uparrow},N_{\downarrow})$ in each sector can be numerically calculated
with straight forward algorithms.

   In Fig.~\ref{Gap.fig} we have plotted the spin
and charge gaps versus $U$ for a fixed value of $\Delta=20$. 
The numerical calculation is done for $L=20$ lattice sites, subject
to periodic boundary conditions.
This figure shows two transition points at $U_{c_{1}}=19.7$
 and $U_{c_{2}}=20$. For $U<U_{c_{1}}$
spin and charge gap are positive and identical. The gaped charge excitations
characterizes an insulating state. However, since the spin gap 
coincides with the charge gap, the resulting insulating state must be a
simple band insulator. For $U_{c_{1}}<U<U_{c_{2}}$ spin and charge
gap are both zero, characteristic of metallic states~\cite{Bouadim,Garg,Sorella,Craco}.

Finally, for $U>U_{c_{2}}$ charge excitations will become gaped,
while spin excitations remain gap-less. Low-energy spin excitations,
with gaped charge excitation is characteristic of Mott insulators.
This picture is in agreement with the one emerging from
Fig.~\ref{ionicity.fig}, according to which,
$n_{i}\equiv\sum_{\sigma}n_{i\sigma}=1$.

\begin{figure}[t]
  \begin{center}
    \vspace{0.0 cm}
    \includegraphics[scale=0.60,angle=-90]{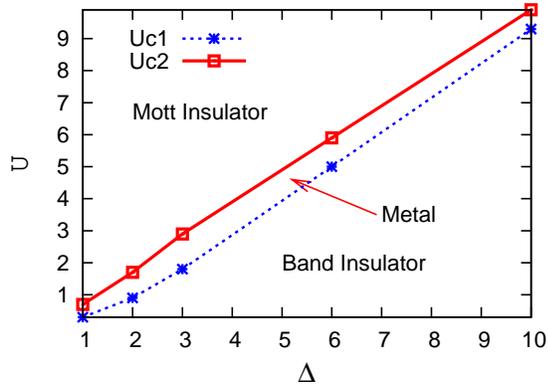}
    \caption{Phase diagram of the IHM base on the effective 
    Hamiltonian. 
    }
    \label{transitionregion.fig}
  \end{center}
\end{figure}

For $\Delta=1$ the
transition points occur in $U_{c_{1}}=0.3$ and $U_{c_{2}}=0.7$ and
for $\Delta=2$ in $U_{c_{1}}=0.9$ and $U_{c_{2}}=1.7$.
The phase diagram depicting the intermediate metallic region is 
shown in Fig.~\ref{transitionregion.fig}. In absence of $t$ term,
the metallic region is characterized by line $ U= \Delta$
In this case both $U_{c_1}$ and $U_{c_2}$ coincide and the metallic
region shrinks to a line. However, inclusion of quantum fluctuations
via $t$ term, induces an attractive $V$ term which eventually broadens
the metallic line $U=\Delta$ into the narrow region shown in Fig.~\ref{transitionregion.fig}.
In the limit $U,\Delta\gg t$ where the effective Hamiltonian reduces 
to the atomic limit~(\ref{atomiclimit.eqn}), the intermediate region for a fixed $\Delta$
shrinks to zero and the metallic remains single point 
at $U_c=\Delta$~\cite{Bouadim}. Therefore the large $\Delta,U\gg t$ limit
of the Fig.~\ref{transitionregion.fig} is a line on which 
band insulating and Mott insulating phases are degenerate.

In the ground state $|\tilde\Psi_0\rangle=|\Psi_0(\infty)\rangle$ 
of~(\ref{Heff.eqn}), the quantum fluctuations are frozen.
The ground state of the original fermions is given by 
$|\Psi_0\rangle=\hat U(\infty)|\tilde\Psi_0\rangle$. 
When this transformation is applied,
it is expected to restore the low-energy spin spectrum of the Mott insulating
state, as well as the excitation spectrum of the resulting metallic state.

The question of whether the intermediate metallic state is a Fermi liquid,
or non-Fermi liquid can be investigated by studying the dynamic correlation
functions $\langle \Psi_0| O_j(t) O_0(0)|\Psi_0\rangle$.
But since it is difficult to calculate $|\Psi_0\rangle$ of the
original fermions, one can use the generator to set up flow equations for the 
observable $\hat O$, to calculate instead the quantity  
$\langle \tilde\Psi_0|\tilde O_j(t) \tilde O_0(0)|\tilde\Psi_0\rangle$.
As far as the calculation of static quantities such as various energy
gaps presented in this paper is concerned, the way one does the
unitary transformation will not be important. But for dynamic
correlation functions in order to capture the incoherent features
in the electronic structure of strongly correlated systems such as
IHM, one needs to organize the hopping processes depending on whether
they change the double occupancy or not~\cite{Uhrig-t-J,KehreinDiscussion}.

  To summarize, we have presented a strong coupling treatment of the
ionic Hubbard model within the CUT approach. This enabled us to map
the IHM into an effective Hamiltonian containing only commuting variables
$n_{i\sigma}$. Getting rid of fermionic minus sing problems in this way,
enables us to investigate the energetics of the model Hamiltonian, suggesting
the phase diagram presented in Fig.~\ref{transitionregion.fig}.

\section{Acknowledgement}
This work was partially supported by ALAVI Group Ltd.
We would like to thank S. Kehrein, F. Shahbazi, M. Moeckel, P. Fritsch, 
for useful discussions. S.A.J. thanks S. Kehrein for hospitality
during his visit to LMU, Munich.


\begin{thebibliography}{50}
\bibitem{Nagaosa} N. Nagaosa and J. Takimoto, J. Phys. Soc. Jpn.
{\bf 55}, 2735 (1986).
\bibitem{Egami} T. Egami, S. Ishihara, and M. Tachiki, Science. {\bf 261}, 1307 (1993).
\bibitem{RefolioJPCM} M. C. Refolio, J. M. Lopez Sancho, and J. Rubio, J. Phys. Condens. Matter, {\bf 17}, 6635 (2005).
\bibitem{Sorella} N. Gidopoulos, S. Sorella, E. Tosatti, E. Phys. J. B {\bf 14}, 217 (2000).
\bibitem{Fazekas} P. Fazekas, {\em Lecture notes on electron correlation and Magnetism}, 
World scientific, 1999
\bibitem{TohyamaNLO} T. Tohyama, S. Maekawa, J. Luminescence, {\bf 94-95}, 659 (2001).
\bibitem{Manmana} S. R. Manmana, V. Meden, R. M. Noack, and K. Sch\"{o}nhammer, Phys. Rev. B {\bf 70}, 155115 (2004).
\bibitem{Legeza} \"{O}. Legeza, K. Buchta, and J. S\'{o}lyom, Phys. Rev. B {\bf 73}, 165124 (2006).
\bibitem{Brune} Ph. Brune, G. I. Japaridze, A. P. Kampf, and M. Sekania, cond-mat/0106007.
\bibitem{Bouadim} K. Bouadim, N. Paris, F. H\'{e}bert, G. G. Batrouni, and R.T.
Scalettar, Phys. Rev. B {\bf 76}, 085112 (2007); 
N. Paris, K. Bouadim, F. H\'{e}bert, G. G. Batrouni, and R. T.
Scalettar, Phys. Rev. Lett. {\bf 98}, 046403 (2007).
\bibitem{Wilkens} T. Wilkens and R.M. Martin, Phys. Rev. B {\bf 63}, 235108 (2001).
\bibitem{Refolio} M. C. Refolio, J. M. Lopez Sancho, and J. Rubio, cond-mat/0210462.
\bibitem{Garg} A. Garg, H. R. Krishnamurthy, and M. Randeria, phys.
Rev. Lett. {\bf 97}, 046403 (2006).
\bibitem{Craco} L. Craco, P. Lombardo, R. Hayn, G.I. Japaridze, and E.
Muller-Hartmann, cond-mat/0703814v2 (2007).
\bibitem{Kancharla} S. S. Kancharla and E. Dagotto, Phys. Rev. Lett. {\bf 98}, 016402 (2007).
\bibitem{Fabrizio} M. Fabrizio, A. O. Gogolin and A. A. Nersesyan, Phys. Rev. Lett. {\bf83}, 2014 (1999).
\bibitem{Batista} C. D. Batista and A. A. Aligia, Phys. Rev. Lett. {\bf 92}, 246405 (2004).
\bibitem{KehreinBook} S. Kehrein, {\em The Flow Equation Approach to Many-Particle Systems}, Springer, 2006.
\bibitem{Wilson} S. D. Glazek and K. G. Wilson, Phys. Rev. D {\bf 48}, 5863 (1993).
\bibitem{Wegner} F. Wegner, Ann. Phys. (Leipzing) {\bf 3}, 77 (1994); 
F. Wegner, J. Phys. A: Math. Gen. {\bf 39}, 8221 (2006).
\bibitem{Kehrein} S. Kehrein and A. Mielke, J. Phys. A {\bf 27}, 4259 (1994).
\bibitem{Stein} J. Stein, J. Phys. G {\bf 26}, 377 (2000).
\bibitem{Lenz} P. Lenz and F. Wegner, Nucl. Phys. B {\bf 482}, 693 (1996).
\bibitem{Sine} S. Kehrein, Phys. Rev. Lett. {\bf 83}, 4914 (1999).
\bibitem{Stein99} J. Stein, Eur. Phys. J. B {\bf12}, 5 (1999).
\bibitem{Moeckel} M. Moeckel and S. Kehrein, Phys. Rev. Lett. {\bf100}, 175702 (2008).
\bibitem{Heidbrink} C. P. Heidbrink, G. S. Uhrig, Phys. Rev. Lett. {\bf 88}, 146401 (2002).
\bibitem{Mielke} A. Mielke, Eur. Phys. J. B {\bf 5}, 605 (1998).
\bibitem{Dziurzik} C. Dziurzik, G. I. Japaridze, A. Schadschneider, and J. Zittartz, Eur. Phys. J. B {\bf 37}, 453463 (2004).
\bibitem{Uhrig-t-J} A. Reischl, E. Muller-Hartmann, G. S. Uhrig, Phys. Rev. B {\bf 70}, 245124 (2004).
\bibitem{KehreinDiscussion} S. Kehrein, private communication
\end{thebibliography}
\end{document}